\documentclass[prb,preprint]{revtex4-1}
\usepackage{amsmath}  
\usepackage{amsfonts} 
\usepackage{graphicx} 
\begin{document}
\title{Average Lorentz Self-Force From Electric Field Lines}
\author{Sandeep Aashish}
\email{ sandeepaashish@gmail.com} 
\author{Asrarul Haque }
\email{ahaque@hyderabad.bits-pilani.ac.in} \altaffiliation[B209,
Department of Physics ]{BITS Pilani Hyderabad campus, Jawahar Nagar,
Shameerpet Mandal, Hyderabad-500078, AP, India.}
\affiliation{Department of Physics, BITS Pilani Hyderabad Campus,
 Hyderabad-500078, AP, India.}
\date{\today}
\begin{abstract}
We generalize the derivation of electromagnetic fields of a charged
particle moving with a constant acceleration\cite{singal} to a
variable acceleration (piecewise constants) over a small finite time
interval using Coulomb's law, relativistic transformations of
electromagnetic fields and Thomson's construction\cite{thmsn}. We
derive the average Lorentz self-force for a charged particle in
arbitrary non-relativistic motion via averaging the fields at
retarded time.
\end{abstract}
\maketitle 
\section{Introduction}
The electromagnetic fields\cite{singal} of a charged particle moving
with a constant acceleration are obtained exploiting Coulomb force,
relativistic transformations of electromagnetic fields and Thomson's
construction\cite{thmsn}. The derivation of the fields for an
accelerated charge is carried out in the instantaneous rest frame.
The geometry of the Thomson's construction makes it evident that
fields pick up the transverse components proportional to
acceleration in addition to the radial components. The
electromagnetic fields so obtained turn out exactly the same as
those obtained via Lienard-Weichert potentials. This method is
mathematically simpler than the usual method\cite{jack1} of
computing the electromagnetic fields which involves rather
cumbersome calculations.\\ However, the one downside to this method
is that it is not sufficient to calculate the radiation reaction
force. The derivation of radiation reaction involves
non-uniform acceleration of the charged particle.\\
In this paper, we address the question as to how to calculate the EM
fields for a charged particle moving with a variable acceleration
(piecewise constants) over a small finite time interval $\Delta t$
using relativistic transformations of fields $\vec E$ and $\vec B$,
Coulomb field of a stationary charge as well as Thomson's
construction. Consider a charge moving with piecewise different
constant accelerations in time interval $\Delta t$. The charge
moving with piecewise $N$(say) different constant accelerations over
time interval $\Delta t$ enables us to use the relativistic
transformations of fields $\vec E$ and $\vec B$ of a uniformly
moving charge through $N$ small time sub-intervals $\Delta
t/N$(say). The electromagnetic fields at some space and time points
are obtained by time-averaging out the fields stemming from the
piecewise $N$ different constantly accelerated motions of the charge
through $N$ temporal sub-intervals $\Delta
t/N$.\\
A charged particle  moving with non-uniform acceleration radiates. A
radiating charged particle experiences a force which acts on the
charge particle and is called as self-force. The Lorentz self-force
\cite[p.~753]{jack1} arising due to a point charge conceived as a
uniformly charged spherical shell of radius $s$ is given by
 \begin{equation}
\vec F_{self}  =  - \frac{2}{3}\frac{{q^2 }}{{4\pi  \varepsilon_0
c^2s }}\dot {\vec v}(t) + \frac{2}{3}\frac{{q^2 }}{{4\pi
\varepsilon_0 c^3 }}\ddot{\vec v}(t)+ O(s)~~\textup{with}~~|\vec s|=
s \label {aa}\end{equation} where,
\begin{itemize}
\item the quantity $\frac{2}{3}\frac{{q^2 }}{{4\pi \varepsilon
_0 c^2s }}$ in the first term stands for electromagnetic mass and
becomes
divergent 
as $s\to 0^+$,
\item the second term represents the radiation reaction and is independent of
the dimension of the charge distribution and
\item the third term corresponds to the first finite size correction and is proportional to the
radius of the shell $s$.
\end{itemize}
It is plausible to expect that the piecewise $N$ different
constantly accelerated motions of the charge through $N$ temporal
sub-intervals $\Delta t/N$ could give rise to the average
self-force. We derive the average Lorentz self-force for the charged
particle in arbitrary non-relativistic motion via averaging the said
retarded fields\cite{haque}.
\section{Preliminary}
In this section, we shall briefly discuss about the relativistic
transformations of the fields and Thomson's
construction\cite{thmsn}. We shall further discuss and summarize the
results on the electromagnetic field of a constantly accelerated
charge in the instantaneous rest frame\cite{singal} and
self-force\cite{haque}.
\subsection{Relativistic Transformations of $\vec E$ and $\vec B$ of a Uniformly Moving Charge}
Let us consider two frames $S$ and $S'$. Let $S'$ is moving with
constant velocity $\vec v = \vec {\beta} c$ relative to $S$. Suppose
a particle of charge $q$ moves with a velocity $\vec v$ relative to
$S$. The charged particle would thus appear to be at rest with
respect to the
system $S'$.\\
The electric $\vec E$ and magnetic fields\cite{djg} $\vec B$ of the
charged particle in frame $S$ is related to the electric $\vec E'$
 and magnetic fields $\vec B'$ of the charged particle in the frame
$S'$ as follows:
\begin{eqnarray}
\vec{E}&=&\vec{E'}_{\parallel}+\gamma[\vec{E'}_{\perp}-\vec
{\beta}\times\vec{B'}]  ~,~
\vec{B}=\vec{B'}_{\parallel}+\gamma[\vec{B'}_{\perp}+ \vec {\beta}\times\vec{E'}]\nonumber \\
\vec{E}^\prime&=&\vec{E}_{\parallel}+\gamma[\vec{E}_{\perp}+\vec
{\beta}\times\vec{B}] ~,~
\vec{B}^\prime=\vec{B}_{\parallel}+\gamma[\vec{B}_{\perp}-\vec
{\beta}\times\vec{E}]
\end{eqnarray}
In case, the charged particle moves with non-relativistic speed
$|\vec {\beta}| <<1~i.e.~ \gamma \to 1$, $\vec E$ and $\vec B$ field
transformations simplify to yield:
\begin{eqnarray}
\vec{E}&=&\vec{E'}-\vec {\beta}\times\vec{B'}  ~,~
\vec{B}=\vec{B'}+ \vec {\beta}\times\vec{E'} \\
\vec{E}^\prime&=&\vec{E}+\vec {\beta}\times\vec{B} ~,~
\vec{B}^\prime=\vec{B}-\vec {\beta}\times\vec{E}
\end{eqnarray}
In $S'$ frame, field is purely electric as the charge is at rest
with respect to the system $S'$. Therefore,
\begin{equation}
\vec{B'}=0, \vec{E}=\vec{E'}~\textup{and}~\vec{B}=\vec
{\beta}\times\vec{E'}
\end{equation}
Now, suppose that the charged particle is moving along the
Z-axis($\theta =0$) i.e. $\vec {\beta}=\beta \hat z$, then in the
spherical polar coordinates $(R,\theta,\phi)$, we have
\begin{eqnarray}
\vec{E}&=& E_R \hat{R} = \frac{e^2}{R^2}\hat{R}\\
\vec{E'}&=& E_R'\hat{R'} =\frac{e^2}{R'^2}\hat{R'}\\
\vec{B}&=& \vec {\beta}\times\vec{E'}= \vec
{\beta}\times\vec{E}=B_{\phi} \hat{\phi}= \frac{e\beta \sin
\theta}{R^2} \hat{\phi}
\end{eqnarray}
Now, $\vec{E}= \vec{E'}$ implies that $R=R'$ where $R$ is the
distance between the field point and the location of the charge in
the frame $S$.
\subsection{Thomson's Construction}
Consider a charged particle, initially moving with a constant
velocity $\vec{v}_I,$ suffers a change in velocity after the time
interval $(0,\tau)$ to a constant velocity $\vec{v}_F$. Suppose the
charged particle undergoes an acceleration to a small velocity
$\Delta \vec v$ ($\Delta v<<c$) for the short time $\tau$.
Arguments due to Thompson, regarding the resulting field
distribution in terms of the electric field lines after time $t =
T$, attached to the accelerated charge are summarized as follows:
\begin{figure}[hbtp!]
\begin{center}
    \includegraphics[bb = 120 240 600 550,
    scale=0.7,angle=0]{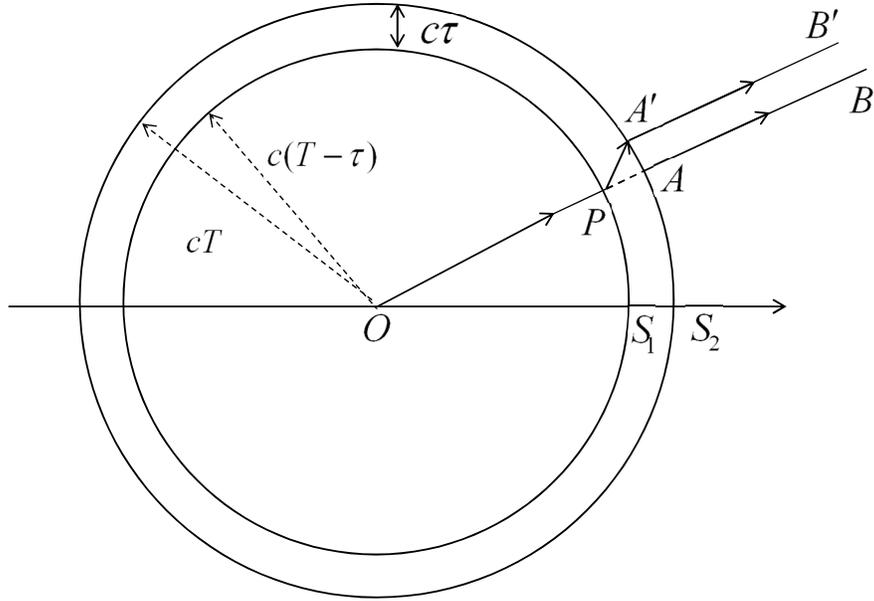}
       \caption{Electric field line as per Thomson's construction exhibiting the 'kink' in between the spherical
       surfaces $S_1$ and $S_2$ corresponding to the acceleration of the charge.}
\label{cas1}
\end{center}
\end{figure}
\begin{itemize}
\item For any time $t<0$,
  fields are that of the charge moving with constant velocity
  $\vec{v}_I$. The electric field lines will emanate radially outward
  from the charge in all possible directions. The information pertaining to the change in
  motion (acceleration)
  can't reach outside a
  sphere of radius $R = cT$.
\item For any time $0< t < \tau$,
  fields are that of the charge undergoing acceleration. The electric field lines will admit distortions
  in the form of a kink in a region between the two spheres $S_1$ and $S_2$ (as shown in FIG.\ref{cas1} )
   in order to preserve the continuity of the field
  lines. Thus, the fields would now begin to pick up the tangential
  component in addition to the radial one.
  The information pertaining to the change in motion
 is confined in the spatial region $cT< R < c(T-\tau)$.
\item For any time $t >\tau$
  fields are that of the charge moving with constant velocity $\vec{v}_F.$ The electric field
  lines will emanate radially outward
  from the charge in all possible directions.
  The information pertaining to the change in motion
  can't reach inside a
  sphere of radius $R= c(T-\tau)$.
\end{itemize}
\subsection{Electromagnetic Field of the Constantly Accelerated Charge\cite{singal}}
 Consider a charged
particle moving in a lab-frame $S$. Suppose the charge is moving
with a constant initial velocity $\vec{v}_{0}$. Let the charge be
uniformly accelerated $\vec{a}$ for a short time interval ($-\Delta
t/2,\Delta t/2$) to a velocity $\Delta \vec v (\Delta v<<c)$ so that
its velocity become $\vec{v}_{1}$ (constant). The velocity of the
charge at $t=0$
is $\vec{v}=(\vec{v}_{0}+\vec{v}_{1})/2$.\\
Consider a (instantaneous rest)
 frame $S'$ moving with
  velocity $\vec{v}=(\vec{v}_{0}+\vec{v}_{1})/2$ relative to frame $S$. The initial and final
  velocities say $\vec{v'}_{0}$ and $\vec{v'}_{1}$ respectively of the charged particle relative to
  frame $S'$
   turn out equal and
   opposite $\vec{v'}_{1}=-\vec{v'}_{0}\equiv \vec{v'}~ (say)$.
    For convenience,
   $S'$ could be rotated (rotation can be undone at the end) so that
   the charge motion is along the horizontal axis. \\
Suppose the charge be instantaneously at rest at $O'$ at $t'=0$. The
charge moves a distance $a'\Delta t'^2/8$ towards $O'$ and then gets
back in duration $\Delta t'$. To the first order in $\Delta t',$
charge could be assumed to be practically at rest at $O'$. Consider
the fields of the charge at time $T'>>\Delta t'$. Let $O'_1$ and
$O'_2$ (for $N=1$) be the positions of the charge at $T'-\Delta
t'/2$ and $T'+\Delta t'/2$ respectively. The electric field in the
regions $c(T'+\Delta t'/2)< r' < c(T'-\Delta t'/2)$ would be in the
radial direction from the points $O'_1$ and $O'_2$. We wish to
calculate the electric field in the region $c(T'-\Delta t'/2)<
\Delta r' < c(T'+\Delta t'/2)$ which possesses the information of
the change in motion of the charge.
\\
Geometrically, it is obvious from the Thomson's construction (for
$N=1$), that the field now picks up both the radial ($E'_{r'}$) as
well as the transverse components ($E'_{\theta '}$) both at $A$ and
$B$. The spatial variation in the transverse components of electric
field over a distance $\Delta r'$ from $A$ to $B$ turns out,
\begin{equation}
\frac{{\partial E'_{\theta '} }}{{\partial r'}} = \frac{{ -
e}}{c}\frac{{\dot \beta '\sin \theta '}}{{r'^2 }}\label{eq:03}
\end{equation}
The formal solution of (\ref{eq:03}) at $P(r',t)$ assuming that
field falls to zero as $r'\to \infty$ leads to
\begin{equation}
E'_{\theta '}  = \frac{{e\dot \beta '\sin \theta '}}{{cr' }}
\end{equation}
The total electric field at $P(r',t)$ could be written as:
\begin{equation}
\vec{E}^\prime=\frac{q}{r'^2}\hat{r'}+\frac{q}{c^2}\frac{\vec{r'}\times(\vec{r'}\times
\dot{\vec{v'}})}{r'}
\end{equation}
The transformation of the field from $S'$ to $S$ yields:
\begin{eqnarray}
\vec{E}=q\frac{\hat{r}-\vec \beta}{r^2 \gamma^2 (1-\hat{r}\cdot \vec
\beta)^3} + \frac{q}{c}\frac{\hat{r}\times\{(\hat{r}-\vec
\beta)\times \dot{\vec \beta}\}}{r(1-\hat{r}\cdot\vec \beta)^3}
\end{eqnarray}
In the non-relativistic case ($\frac {v}{c}<<1~i.e.~\gamma
\rightarrow 1$), the expression for $\vec{E}$ takes the form:
\begin{eqnarray}
\vec{E}=q\frac{\hat{r}-\vec \beta}{r^2 (1-\hat{r}\cdot \vec
\beta)^3} + \frac{q}{c}\frac{\hat{r}\times\{(\hat{r}-\vec
\beta)\times \dot{\vec \beta}\}}{r(1-\hat{r}\cdot\vec \beta)^3}
\end{eqnarray}
\subsection{Results of The Self-force\cite{haque}}
 A simple derivation of the
self-force\cite{haque} based on the consideration that the averaged
value of the field in the suitably small closed region surrounding
the point charge is the value of the field under
consideration at the position of the point charge is carried out in detail.\\
The self-force is defined as:
\begin{equation}
\vec F_{Self}(\vec r,t) = q\mathop {Lim}\limits_{s \to 0^+}
\overline {\vec E}(\vec r,t) = q\vec E_{Self}(\vec r,t)
\end{equation}
where $\overline {\vec E}(\vec r,t))$ is average field over the
surface of a spherical shell of radius $s$ and the field ${\vec
E}(\vec r,t)$ depends upon the position and motion of the charge
particle at the retarded time. Using the field due to an accelerated
charged particle (in the limit $v/c \to 0 $), the self force turns
out:
\begin{equation}
\vec F_{Self}(\vec r,t)= -\frac{2}{3}\frac{q^2}{4\pi
\epsilon_{0}c^{2}}\left(\mathop {Lim}\limits_{s \to 0^+}\frac{\vec
a(t-s/c)}{s}\right)
\end{equation}
\section{Calculation of Average Electric Field}
\begin{figure}[hbtp!]
\begin{center}
    \includegraphics[bb = 120 80 600 550,
    scale=0.7,angle=0]{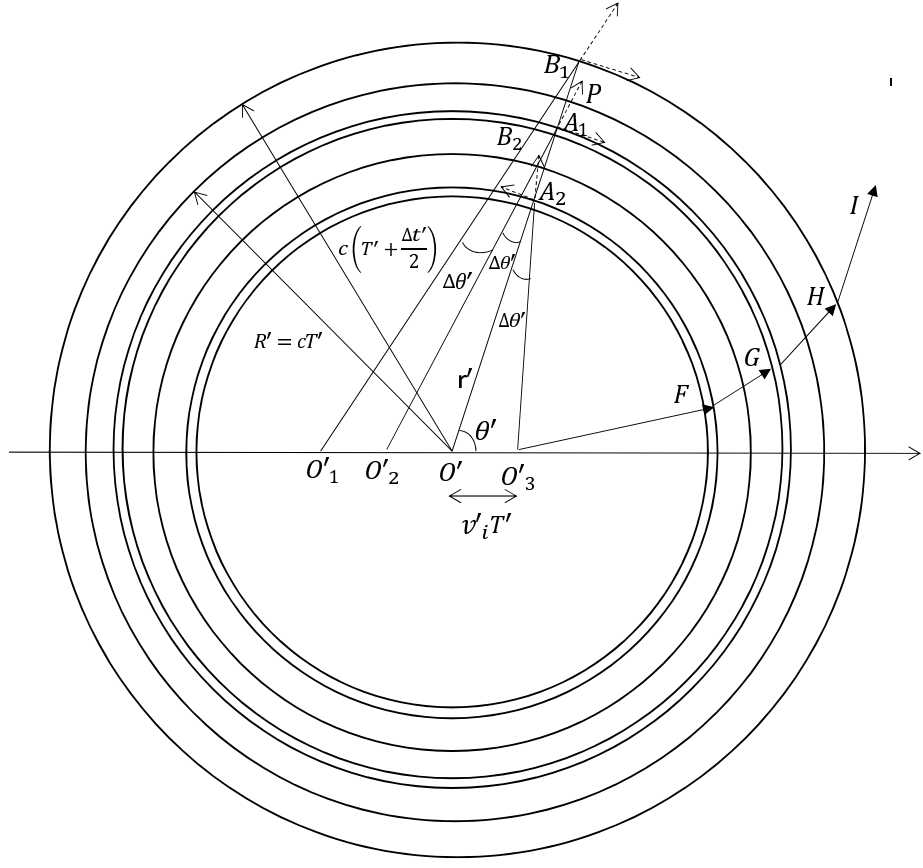}
       \caption{Electric field line $O_3FGHI$ as per Thomson's construction exhibiting the 'kink' in between
       the innermost and the outermost spherical
       surfaces corresponding to the accelerated motion of the charge. The thin annular region
        between the spherical surfaces belongs to the non-accelerated motion of the charge.}
\label{ca04}
\end{center}
\end{figure}

Consider a charge moving with an initial velocity $\vec{v}_{0}$ in
the lab frame $S$. Suppose it undergoes accelerations from $-\Delta
t/2$ to $\Delta t/2$. We consider that the acceleration in the time
interval $\Delta t$ is not continuous but rather consists of a
series of a finite number of different piecewise constant
accelerations.
 Let us divide the total time interval $\Delta t$ into a large number
 $2N(N\ge2)$ of sub-intervals. Suppose all the odd and even sub-intervals
 are of lengths $(1-\varepsilon)\Delta t/N$ and $\varepsilon\Delta
 t/N$ respectively.
 We consider that the charge undergoes nonzero constant
 accelerations in the odd sub-intervals accompanied by nonzero constant velocities
 in the even sub-intervals.
 Accelerations \{$\vec{a_i}(\tau_i):i=1,3,5,...,2N-1$\} and
  velocities \{$\vec{v}_i(t):i=0,2,...,2N$\} are defined as:\\
\begin{eqnarray}
\vec a_i (\tau _i ) &=& \left\{ \begin{array}{l}
 \vec a_i \theta (\tau _i  - t_{i-1} )\theta (t_{i}-\tau _i )~;~i=1,3,5,...,2N-1 \\
 0,{\rm{  }}(t_{i}  < \tau _i  < t_{i+1} )~;~i=0,2,...,2N \\
 \end{array} \right.\\
\vec{v}_{i+1}(t)&=& \vec{v}_{i}+\int^{\frac{\Delta
t}{2}}_{-\frac{\Delta t}{2}}\vec
a_{i+1}(\tau_i)d\tau_i~;~i=0,2,...,2N
\end{eqnarray}
where,
\[
t_i  = \left\{ \begin{array}{l}
 \left( { - \frac{1}{2} + \frac{i}{{2N}}} \right)\Delta t~;~i=0,2,...,2N. \\
 \left( { - \frac{1}{2} + \frac{{i + 1 - 2\varepsilon }}{{2N}}} \right)\Delta t~;~i=1,3,5,...,2N-1. \\
 \end{array} \right.
\]
\begin{figure}[hbtp!]
\begin{center}
    \includegraphics[bb = 200 225 500 550,
    scale=0.7,angle=0]{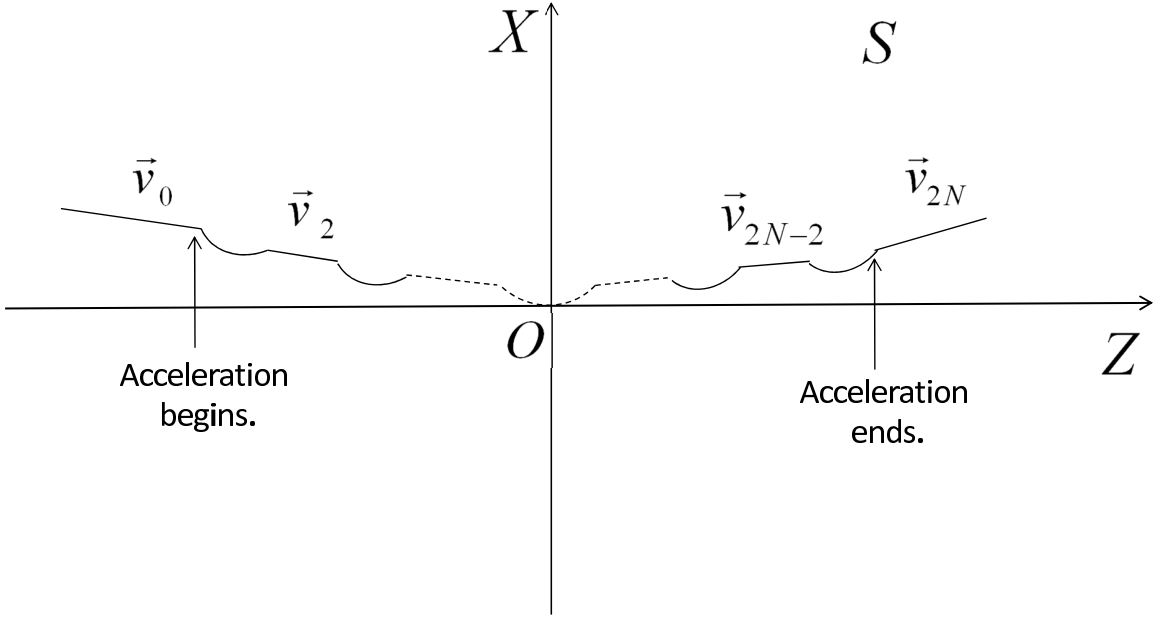}
       \caption{A schematic of the motion of a piecewise constantly accelerated charge.}
\label{casa9}
\end{center}
\end{figure}
 The charge undergoes through various different piecewise constant
 accelerations in the time interval $\Delta t$. In order to determine the EM fields of
 an accelerated charge over $\Delta t$, we require as
 many instantaneous rest-frames
 as that of the different constant accelerations. These
 instantaneous rest-frames could be obtained by appropriate
 boosts.
Let us consider the ith instantaneous rest-frame
\{$S^\prime_{i}:i=1,3,5,...,2N-1$\} moving with velocity
 \{$\vec{v}_{S'_{i}}=(\vec{v}_{i}+\vec{v}_{i+1})/2:i=1,3,5,...,2N-1$\}.
We assume that $|\vec{v}_{i+1}-\vec{v}_{i}|<<c$. Suppose the charged
particle appears to be instantaneuosly at rest at
 $(\frac{t'_i+t'_{i+1}}{2};~
 i=0,2,...,2N-2)$ where
\[
t'_i  = \left\{ \begin{array}{l}
 \left( { - \frac{1}{2} + \frac{i}{{2N}}} \right)\Delta t'~;~i=0,2,...,2N. \\
 \left( { - \frac{1}{2} + \frac{{i + 1 - 2\varepsilon }}{{2N}}} \right)\Delta t'~;~i=1,3,5,...,2N-1. \\
 \end{array} \right.
\]
Thomson's construction for $N=2$ is shown in FIG.\ref{ca04}.
In the frame $S^\prime_{i}$, the corresponding transformed
velocities $\vec{v}^\prime_{i-1}$
 and $\vec{v}^\prime_{i+1}$ are given by:
\begin{eqnarray}
\vec{v}^\prime_{i+1}=-\vec{v}^\prime_{i-1}=
\vec{v}^\prime_{i}~(\textup{say})
\end{eqnarray}
\begin{figure}[hbtp!]
\begin{center}
    \includegraphics[bb = 120 225 500 550,
    scale=0.7,angle=0]{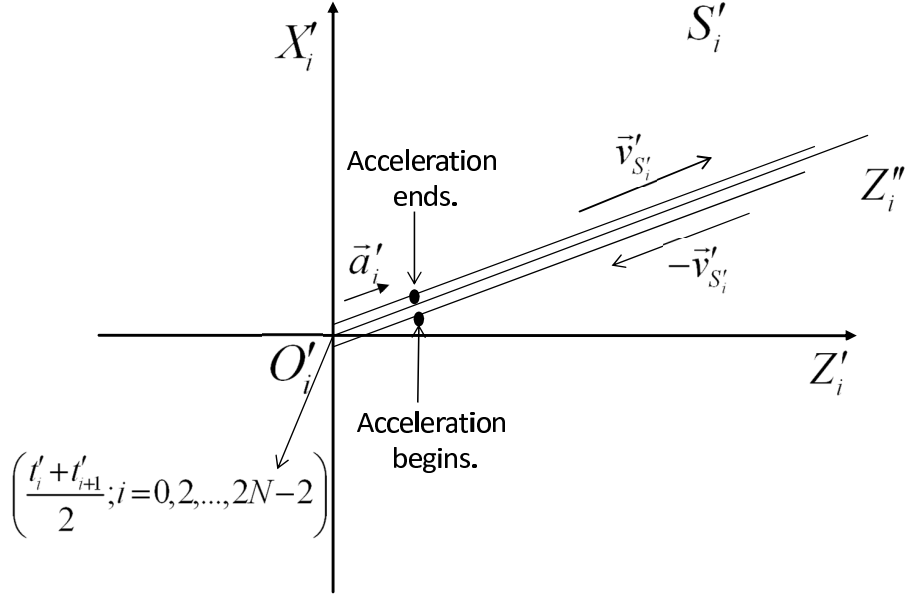}
       \caption{Motion of a piecewise constantly accelerated
        charge as observed in the instantaneous rest frames $S'_i$.}
\label{casa9}
\end{center}
\end{figure}
The acceleration in $S^\prime_{i}$ reads:
\begin{equation}
\vec{a}^\prime_{i}=\frac{2\vec{v}^\prime_{i}}{(1-\varepsilon)\Delta
t'/N}
\end{equation}
Without any loss of generality, we take the orientation of
$S^\prime_{i}$ such that motion happens along $Z'_{i}$. The
calculation of $E^\prime_{i}$ in $S^\prime_{i}$ proceeds in a
similar way to that in the
 section 3.
The electric field $E^\prime_{i}$ at a later time $T'>>\Delta t'/N$
at $P$ is obtained as:
\[
\vec{E}^\prime_{i} =
 \frac{q}{{r'^2 }}\hat n' + \frac{q}{c}\frac{{\hat n' \times
 \hat n' \times \dot {\vec \beta}^\prime_{i} }}{r}
\]
In the above expression, we have made use of
\[ O'_{i}B_{i}\approx
O'_{i+1}A_{i}\approx O'P= r';~i=1,2,...,N \] as
$\Delta{\theta^\prime }$ is small.
Transformation of $\vec{E}^\prime_{i}$ to the lab frame $S$ for the
non-relativistic velocity ($\vec \beta _{i} \to 0,\gamma \to 1 $)
yields:
\[
\vec E_{i}  =
 \frac{q}{{r^2 }}\hat n + \frac{q}{c}\frac{{\hat n \times \hat n \times \dot {\vec \beta} _{i} }}{r}
\]
where, $\hat{n}=\frac{\vec{r}}{r}$ and $ \dot{\vec \beta}_{i}=
\frac{{\vec a_{i}\left (\tau _{i}\right )}}{c}$. The quantities
$\vec r$ and $ \dot{\vec\beta}_{i}$ on the right hand side are
evaluated
 at retarded time, $t_{R_{i}}=\tau _{i} - r/c$.
 The electric field at $P$
\[
\vec E_i  = \left\{ \begin{array}{l}
 \frac{q}{{r^2 }}\hat n + \frac{q}{c}\frac{{\hat n \times \hat n \times \dot {\vec \beta} _i }}{r};~i = 1,3,...,2N-1 \\
 \frac{q}{{r^2 }}\hat n;~i = 0,2,...,2N. \\
 \end{array} \right.
\]
in fact
  consists of $N$ number of piecewise different values of $\vec E_i $ belonging to $2N-1$ sub-intervals
  over
  $\Delta t$.
  It is therefore plausible to consider the electric field at P in the vicinity of a point in time
  as the time-averaged value $\left\langle {\vec E} \right\rangle$
  of the electric fields $\vec E_i $ for the entire time $\Delta t$.
   The time-averaged electric field is obtained as:
\begin{eqnarray}
 \left\langle {\vec E} \right\rangle &=& \frac{{\int\limits_{ - \frac{{\Delta t}}
 {2}}^{\frac{{\Delta t}}{2}} {\sum\limits_{i = 1}^N {\vec E_i dt} } }}
 {{\int\limits_{ - \frac{{\Delta t}}{2}}^{\frac{{\Delta t}}{2}} {dt} }}
  = \frac{{\sum\limits_{i = 1}^N {\left( {\vec E_{2i - 1} \frac{{1 -
  \varepsilon }}{N}\Delta t + \vec E_{2i} \frac{\varepsilon }{N}\Delta t} \right)} }}{{\Delta t}} \nonumber\\
  &=& \frac{1}{N}\sum\limits_{i = 1}^N {\left( {\frac{q}{{r^2 }}\hat n +
  \frac{q}{c}\frac{{\hat n \times \hat n \times \dot{ \vec \beta }_{2i - 1} }}
  {r}} \right)}  - \frac{\varepsilon }{N}\sum\limits_{i = 1}^N {\frac{q}{c}
  \frac{{\hat n \times \hat n \times \dot {\vec \beta }_{2i - 1} }}{r}}
\end{eqnarray}
For $\varepsilon << \frac{\Delta t}{N}$, we have:
\begin{equation}
\left\langle {\vec E} \right\rangle  = \frac{q}{{r^2 }}\hat n +
\frac{q}{c}\frac{{\hat n \times \hat n \times \left\langle {\dot
{\vec \beta } } \right\rangle }}{r}
\end{equation}
where,
\begin{equation}
\left\langle {\dot {\vec \beta}} \right\rangle  =
\frac{{\frac{1}{c}\int\limits_{ - \frac{{\Delta
t}}{2}}^{\frac{{\Delta t}}{2}} {\vec a_{2i - 1} (\tau _{2i - 1} )}
d\tau _{2i - 1} }}{{\Delta t}} =
   \frac{{\sum\limits_{i = 1}^N {\frac{{\vec a_{2i - 1} (\tau _{2i - 1} )}}{c}} }}{N} 
\end{equation}
\section{Calculation of the Self force}
The self force of a charge moving with arbitrary velocity, in
general, contains acceleration and higher derivatives of
acceleration as is especially obvious from equation (\ref{aa}). A
charge moving with constant acceleration does not experience any
 radiation reaction as the term
 $\frac{2}{3}\frac{q^{2}}{4\pi\epsilon_{0}c^3}\ddot{\vec{v}}(t)$ vanishes.
\\
In the case at hand, charge is moving with non-zero constant
acceleration in the time interval $(1-\varepsilon)\Delta t/N$
whereas with zero acceleration in the time interval
$\varepsilon\Delta t/N$. Therefore, the charge confined to these
time intervals will not experience any radiation reaction force.
However, over the interval $\Delta t$, the charge moving with
various different
 constant accelerations would give rise to a net change in the acceleration
 $\Delta \vec{a}$ over $\Delta t$.
  This suggests that over
 the time $\Delta t$, $\Delta \vec{a}/\Delta t$ is no longer zero, and hence the  charge
  must experience average radiation reaction. The average
  self-force\cite{haque} may be defined as
\begin{equation}
\vec F_{Self} (\vec r,t) =
 - \frac{2}{3}\frac{{q^2 }}{{4\pi \varepsilon _0 c^2 }}
 \mathop {Lim}\limits_{s \to 0^ +  } \frac{{\left\langle
  {\vec a(\tau _{2i - 1}  - s/c)} \right\rangle }}{s}
\end{equation}
where,
\[
\vec a_{2i - 1} (\tau _{2i - 1}  - s/c) = \vec a_{2i - 1} \theta
(\tau _{2i - 1}  - t_{2i - 2}  - s/c)\theta (t_{2i - 1}  - \tau _{2i
- 1}  + s/c)
\]
We can Taylor expand $\vec a_{2i - 1} (\tau _{2i - 1}  - s/c) $
about $s/c$ so that,
\[
\vec a_{2i - 1} (\tau _{2i - 1}  - s/c) =\vec a_{2i - 1} (\tau _{2i
- 1})-\frac{s}{c}\frac{d}{d\tau _{2i - 1}}\vec a_{2i - 1}(\tau _{2i
- 1})+ O(s^2)
\]
The self force expression now becomes
\[
\vec F_{Self} (\vec r,t) =  - \frac{2}{3}\frac{{q^2 }}{{4\pi
\varepsilon _0 c^2 }}\mathop {Lim}\limits_{s \to 0^ +  }
  \frac{
\left\langle {\vec a(\tau _{2i - 1} )} \right\rangle }{s} +
\frac{2}{3}
  \frac{{q^2 }}{{4\pi \varepsilon _0 c^3 }}
\left\langle {\dot \vec a(\tau _{2i - 1} )} \right\rangle
\]
 Since,
\begin{eqnarray*}
 \theta (\tau _{2i - 1}  - t_{2i - 2}  - s/c) = \theta (\tau _{2i - 1}
   - t_{2i - 2} ) - \frac{s}{c}\delta (\tau _{2i - 1}  - t_{2i - 2} ) + O(s^2 ) \\
 \theta (t_{2i - 1}  - \tau _{2i - 1}  + s/c) = \theta (t_{2i - 1}  - \tau _{2i - 1} )
  + \frac{s}{c}\delta (t_{2i - 1}  - \tau _{2i - 1} )
 \end{eqnarray*}
Therefore,
\begin{eqnarray*}
&&
 \sum\limits_{i = 1}^N {\vec a_{2i - 1} (\tau _{2i - 1}  - s/c)}  = \sum\limits_{i = 1}^N
 {\vec a_{2i - 1} \theta (\tau _{2i - 1}  - t_{2i - 2} )\theta (t_{2i - 1}  - \tau _{2i - 1} )}  \\
 && + \frac{s}{c}\sum\limits_{i = 1}^N {\vec a_{2i - 1} } \theta (t_{2i-1}  - t_{2i - 2} )
  [\delta (t_{2i - 1}  - \tau _{2i - 1} ) - \delta (\tau _{2i - 1}  - t_{2i - 2} )] + O(s^2 )
 \end{eqnarray*}
It is evident that $ \vec a_{2i - 1} (\tau _{2i - 1}  - s/c)$ turns
out divergent at the temporal boundaries:
\[\tau _{2i - 1}= t_{2i - 1} ~ ~\textup{or}~~\tau
_{2i - 1} = t_{2i - 2}.\] However, the time-averaged self force
would render $\vec a_{2i - 1} (\tau _{2i - 1}  - s/c)$ physically
sensible. Moreover, in order to prevent this meaningless results, we
assume that
 the transitions from non-zero constant acceleration to zero constant
 acceleration are smooth at the temporal boundaries (please see Appendix \ref{apx1} for clarification).
 Such sort of meaningless
results arise in models that involve step functions
forces\cite{sch}. Now,
 \begin{eqnarray}
 \vec F_{Self} (\vec r,t)  &=  &-
\frac{2}{3}\frac{{q^2 }}{{4\pi \varepsilon _0 c^2 s}}\left\langle
{{\vec a}}\right\rangle\nonumber\\
&-&\frac{2}{3}\frac{{q^2 }}{{4\pi \varepsilon _0 c^3}}
\frac{1}{{\Delta t}}\int\limits_{ - \frac{{\Delta
t}}{2}}^{\frac{{\Delta t}}{2}} {\sum\limits_{i = 1}^N {\vec a_{2i -
1} [\delta (t_{2i - 1}  - \tau _{2i - 1} ) - \delta (\tau _{2i - 1}
- t_{2i - 2} )]d\tau _{2i - 1} } }\nonumber\\
&=& - \frac{2}{3}\frac{{q^2 }}{{4\pi \varepsilon _0 c^2
s}}\left\langle {{\vec a}}
  \right\rangle  + \frac{2}{3}\frac{{q^2 }}{{4\pi \varepsilon _0 c^3 }}\frac{1}{2\pi}\left\langle
   {\dot {\vec a}} \right\rangle
  \end{eqnarray}
where we have identified \[ \left\langle {\dot {\vec a}}
\right\rangle = \frac{\vec a_{2i - 1}-\vec a_1}{\Delta t}.
\]
Thus, the time-average radiation reaction stems from the
time-averaged acceleration.
\section{Conclusion}
We derive the electromagnetic fields of a charged particle moving
with a variable acceleration (piecewise constants) over a small
finite time interval using Coulomb's law, relativistic
transformations of fields and Thomson's construction. We derive the
expression for the average Lorentz self-force for a charged particle
in arbitrary non-relativistic motion via averaging the retarded
fields.
\appendix   
\section{A model for piecewise constant and smooth acceleration}\label{apx1}
In order to have the physically sensible values of $ \vec a_{2i - 1}
(\tau _{2i - 1}  - s/c)$, we assume that
 the transition from non-zero constant
acceleration to zero
 acceleration and viceversa is smooth at the temporal boundaries. We can incorporate the smooth change
 in the acceleration at the temporal boundaries by defining our acceleration as follows:
\[
\vec a(\tau _i ) = \left\{ \begin{array}{l}
 \vec a_{2i - 1} ,t_{2i - 2}  + \varepsilon _1  \le \tau _i  \le t_{2i - 1}  - \varepsilon _1 {\rm{( }}i = 1,2,...,N .{\rm{ )}} \\
 \vec a_{2i - 1}  - \frac{{\vec a_{2i - 1} }}{{2\varepsilon _1 }}(\tau _i  - t_{2i - 1}  + \varepsilon _1 ),t_{2i - 1}  - \varepsilon _1  \le \tau _i  \le t_{2i - 1}  + \varepsilon _1 (i = 1,2,...,N - 1.){\rm{ }} \\
 \frac{{\vec a_{2i - 1} }}{{2\varepsilon _1 }}(\tau _i  - t_{2i - 1}  + \varepsilon _1 ),t_{2i - 2}  - \varepsilon _1  \le \tau _i  \le t_{2i - 2}  + \varepsilon _1 (i = 2,3,...,N) \\
 \end{array} \right.
\]
\begin{figure}[hbtp!]
\begin{center}
    \includegraphics[bb = 200 300 600 600,
    scale=0.7,angle=0]{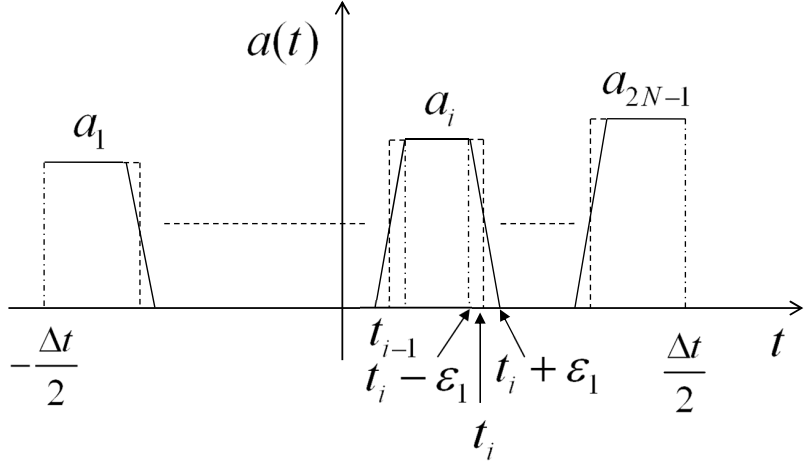}
       \caption{Piecewise constant accelerations having smooth change at the temporal boundaries.}
\label{ca111}
\end{center}
\end{figure}
We assume that $2\varepsilon _1 >> \tau_0,$ where $\tau_0$ (see
Ref.\cite{moni}) is defined by
\[\tau_0 = \frac{2}{3} \frac{q^2}{mc^3}.\]
Now,
\[
\dot{\vec a}(\tau _i ) = \left\{ \begin{array}{l}
   - \frac{{\vec a_{2i - 1} }}{{2\varepsilon _1 }},t_{2i - 1}  - \varepsilon _1  \le \tau _i  \le t_{2i - 1}  + \varepsilon _1 (i = 1,2,...,N - 1){\rm{ }} \\
 \frac{{\vec a_{2i - 1} }}{{2\varepsilon _1 }},t_{2i - 2}  - \varepsilon _1  \le \tau _i  \le t_{2i - 2}  + \varepsilon _1 (i = 2,3,...,N) \\
 \end{array} \right.
\]
 The quantity $\left\langle {\dot {\vec a}} \right\rangle $ now turns out:
\begin{eqnarray}
 \left\langle {\dot {\vec a}} \right\rangle &=& \frac{1}{{\Delta t}}\int\limits_{
 - \frac{{\Delta t}}{2}}^{\frac{{\Delta t}}{2}} {\sum\limits_{i = 1}^N {\dot{\vec
 a}({\tau_i})}
   d\tau _{i}}  \\
& = &\frac{{\vec a_{2N - 1}  - \vec a_1 }}{{\Delta t}}
  \end{eqnarray}

\end{document}